\documentclass[aps,prd,groupedaddress,twocolumn,nofootinbib]{revtex4-2}
\usepackage{amssymb,amsmath}
\usepackage{graphicx}
\usepackage{dcolumn}
\usepackage{bm}
\usepackage{tabularx}
\usepackage{color}
\usepackage[colorlinks=true
,urlcolor=blue
,citecolor=blue
,linkcolor=blue
,pagecolor=blue
,linktocpage=true
,pdfproducer=medialab
]{hyperref}
\usepackage[normalem]{ulem}
\usepackage{appendix}
\usepackage{float}
\newcommand{\rt}[1]{\textcolor{red}{#1}}

\begin{document}

\title{LHC and dark matter implications of $t\text{-}b\text{-}\tau$ Yukawa unification in split SUSY GUTs}


\author{Mureed Hussain}
\email{mhussain617@gmail.com}
\affiliation{Higher Education Department Punjab, New Anarkali Road, Shahrah-e-Quaid-e-Azam, Lahore 54000, Pakistan}
\author{Rizwan Khalid}
\email{rizwan\_khalid@lums.edu.pk}
\affiliation{School of Science and Engineering, Lahore University of Management Sciences (LUMS), Opposite Sector U, D.H.A, Lahore 54792, Pakistan}
\author{Cem Salih Un}
\email{cemsalihun@uludag.edu.tr}
\affiliation{Department of Physics, Bursa Uludag˜ University, TR16059 Bursa, Turkey}

\begin{abstract}
	{We investigate a grand unification (GUT) inspired version of the minimal supersymmetric standard model (MSSM)
    based on a left-right symmetric $4$-$2$-$2$ gauge group, incorporating Yukawa coupling unification and current 
    phenomenological constraints. Utilizing a split soft supersymmetry-breaking (SSB) parameter space motivated by 
    flavor symmetries, we analyze the implications of recent results from ATLAS, CMS, LHCb, and dark matter direct 
    detection experiments. Our numerical scans, conducted with SARAH and SPheno, identify viable low-energy regions 
    consistent with third-generation Yukawa unification, the observed Higgs boson mass, dark matter relic density, 
    and flavor observables such as $B \to X_s \gamma$, $B_s \to \mu^+ \mu^-$ and $B_u \to \tau \nu_{\tau} $.
    Our findings suggest that while current bounds severely constrain much of the 
    MSSM-like parameter space, substantial regions remain experimentally viable and testable in the ongoing LHC run 
    and next-generation dark matter experiments.}
\end{abstract}

\maketitle

\section{Introduction}
The Large Hadron Collider (LHC) has established the Standard Model of particle physics (SM)~\cite{Weinberg:1967tq,Salam:1964ry} as an excellent effective field theory at the $O(10 {\rm TeV})$ scale with the experimental program of the ATLAS~\cite{ATLAS:2008xda}, CMS~\cite{Chatrchyan:2008aa} and LHCb~\cite{LHCb:2008vvz} collaborations. These experiments are also at the forefront of probing physics 
beyond the SM (BSM) and have managed to rule out low-scale manifestations of popular BSM scenarios. Within the context of this manuscript, the ATLAS and the CMS experiments have provided stringent lower mass bounds on `new' particles~\cite{ATLAS:2019lff,ATLAS:2024fub,CMS:2021cox,ATLAS:2022rme,CMS:2023zuu,CMS:2023mny,ATLAS:2021yij,CMS:2021beq} suggested by the minimal supersymmetric standard model (MSSM). Likewise, the LHCb provides an indirect check on BSM physics by probing rare decays of B hadrons and 
provide increasingly precise bounds on, for example, the branching ratios $B \rightarrow X_s \gamma$, $B_s \rightarrow \mu^+ \mu^-$ and $B_u \rightarrow \tau \nu_{\tau}$. 
The standard model of cosmology predicts dark matter (DM) as an important constituent in the energy budget ({$27\% $}) of the Universe~\cite{Planck:2018vyg}. Experiments have, over the past couple of decades, explored the possibility of dark matter particles via their direct couplings with 
nuclei\cite{MarrodanUndagoitia:2015veg,DAMIC:2011khz,Gerbier:2014jwa,Bertoni:2013tua,Drukier:2012hj,XENON:2015gkh,Cushman:2013zza,Fatemighomi:2016ree,DARWIN:2016hyl} or in indirect searches \cite{Goodman:2010ku,Beltran:2010ww,Fox:2011pm,Buchmueller:2014yoa,Abercrombie:2015wmb,IceCube:2012ugg,Klasen:2015uma}. In particular, the null result of DM searches has also helped constrain the parameter space 
of various models beyond the SM. 

In this manuscript we consider the effect of the latest bounds from the ATLAS, CMS, and LHCb experiments as well as the dark matter searches to explore the low scale implications and confront a grand unification (GUT) version of the minimal supersymmetric standard model (MSSM) with the current experimental results. 
In particular, we address the question of $t\text{-}b\text{-}\tau$ third generation Yukawa coupling unification consistent with dark matter and LHC constraints and focus on the 
parameter space that can potentially be tested in the current run at the LHC and dark matter search experiments. 
To be precise, we have used a left-right symmetric model based on the $SU(4)_c$ $\times$ $SU(2)_L$ $\times$ $SU(2)_R$~\cite{Pati:1974yy,Melfo:2003xi}
gauge group to feed our GUT scale boundary conditions on the soft supersymmetry breaking (SSB) parameters 
within the framework of gravity mediated SUSY breaking~\cite{Nath:1982zq,Poppitz:1996fw}.

 Within the context of $SU(4)_c$ $\times$ $SU(2)_L$ $\times$ $SU(2)_R$ inspired MSSM models, several recent studies~\cite{Ellis:2020jfc,Shafi:2023ksr} have focused on spectroscopy as we do in this manuscript. However, 
while these papers are recent, there are updated results from 
the LHC both in terms of direct mass bounds on SUSY particles and on branching ratios of rare $B$ decays that these studies do not incorporate. More importantly, we have addressed the 
question of Yukawa coupling unification consistent with dark matter and various LHC constraints. Further, we have been able to find interesting solutions with a more restrictive parameter set.

We underscore the fact that 
whereas a significant portion of the low-energy parameter space in MSSM-like models has been explored (and excluded) by the LHC~\cite{Ellis:2022emx,Han:2016gvr,CMS:2011bul}, there still 
is quite interesting low-scale phenomenology that is consistent with current bounds. 
Therefore, 
it is interesting to ask what kind of minimalist GUT inspired scenarios can provide a low scale 
supersymmetric phenomenology, all the while remaining consistent with the latest bounds from direct 
SUSY and dark matter searches, along with the theoretically well motivated 
constraints arising from requiring Yukawa coupling unification.

In Section~\ref{sec2}, we provide the details of the model that we have employed in the current analysis as 
well as the experimental constraints that we have used. Following this, we present our results in Section~\ref{Results}, 
and discuss the phenomenology relevant to the parameter space consistent with experimental constraints and theoretical requirements. 
We finally present our conclusions in Section~\ref{conclusion}. 

\section{Model parameters and phenomenological constraints}\label{sec2}
One of the markers of many GUT models is Yukawa unification, which can be realized in models based on $E_{6}$ and $SO(10)$ symmetries. Such symmetries have a variety of representations in which the MSSM superfields can reside together with many other fields which play significant roles in the symmetry breaking and alter the low scale implications (see, for instance, \cite{Babu:1987kp,Huitu:1997rr,Zhang:2008jm,Mohapatra:1995xd,Aulakh:1997ba,Babu:2008ep,Chatterjee:2018gca}). In our work we follow a simple breaking mechanism in which $SO(10)$ breaks to its maximal subgroup $SU(4)_c$ $\times$ $SU(2)_L$ $\times$ $SU(2)_R$ (hereafter, for short, $4$-$2$-$2$)~\cite{Lazarides:1980cc,Pati:1974yy,Melfo:2003xi} which in addition to predicting Yukawa coupling unification~\cite{Gogoladze:2011db,Gogoladze:2009bn} also allows for non-universal soft
SUSY breaking gaugino mass terms~\cite{Gogoladze:2009ug,Hussain:2018xiy,Gogoladze:2012ii,Gogoladze:2011aa,Okada:2011wd}. In particular, the SM gauge structure can be produced, for example, from a left-right symmetric model via the breaking mechanism,
\begin{equation*}
        SU (2)_L \times SU (2)_R \times U (1)_X \rightarrow SU (2)_L \times U (1)_Y.
\end{equation*}
If the $U(1)_X$ is the robust $B-L$ symmetry, these models also provide a natural explanation for the conservation of the difference $B-L$, which has consequences for understanding the origin of the observed matter/antimatter asymmetry in the Universe~\cite{VanDong:2020nwb,Ahmed:2020lua}. Charge quantization can be explained if $U (1)_{B-L}$ is formed by a $SU (4)_c$ group using the fact that $SU (3)$ $\times$ $U (1)$ $\subset$ $SU (4)_c$. In our work we follow the following breaking pattern,

\begin{align*}
        SU (4)c \times SU (2)_L \times SU (2)_R \rightarrow  \\
        SU (3)_c \times SU (2)_L \times SU (2)_R \times U (1)_{B-L} \rightarrow\\
        SU (3)_c \times SU (2)_L \times U (1)_Y.
\end{align*}

This breaking mechanism can be realized if $SO(10)$ breaks into $4$-$2$-$2$ via the 
54 dimensional Higgs $54_H$ acquiring a vacuum expectation value (VEV) in the $4$-$2$-$2$ singlet direction (1,1,1), and $4$-$2$-$2$ subsequently breaks into 
$SU (3)_c \times SU (2)_L \times SU (2)_R \times U (1)_{B-L}$ by the 210 dimensional 
Higgs acquiring a VEV. The conservation of $R-$parity depends on the breaking mechanism of $U(1)_{B-L}$ and can be arranged if the 126 dimensional representations 
$126_{H}$ and $\overline{126}_{H}$ are used. The fields in these representations do not have direct Yukawa coupling with the MSSM matter superfields, except those in $126_{H}$; thus they do not complicate the Yukawa interactions. However, one can also include more Higgs representations to accommodate the consistent masses for the SM fermions \cite{Babu:2016bmy}. Such a set up usually breaks the Yukawa unification (YU). On the other hand, if one assumes that the MSSM Higgs fields solely reside in $10_{H}$ and the other Higgs fields residing in higher dimensional representations negligibly interact with the third family fermions, YU can be maintained for the third-family \cite{Joshipura:2012sr}. 
We assume that the various steps in symmetry breaking occur at around the 
scale $M_{GUT}$. It is important to note, however, that the fields residing in the (1,2,2) in decomposition of the 10 dimensional representation of $SO(10)$ do not decouple. Indeed, the fields in this representation further decompose into the $H_{u}$ and $H_{d}$ in order to accommodate a consistent electroweak symmetry breaking pattern~\cite{Babu:2015bna}. We also note that as a consequence of our assumption about symmetry breaking at $M_{GUT}$, the 
$U(1)_{B-L}$ gauge boson and associated gaugino are also very heavy and decouple from the spectrum. In summary, the effective spectrum below $M_{{\rm GUT}}$ is populated by the MSSM particles.

The left-right symmetry and anomaly cancellation for $U(1)_{B-L}$ does require right-handed neutrinos in the spectrum. They can, in principle, interact via Yukawa interactions with $H_{u}$, but tiny masses of left-handed neutrinos require the relevant Yukawa coupling to be of the order of about $10^{-3}$ if the right-handed neutrinos are quite heavy \cite{Abbas:2007ag} unless one imposes inverse seesaw mechanism \cite{Mohapatra:1986bd,Gonzalez-Garcia:1988okv,Khalil:2010iu}. In this case, the MSSM phenomenology remains almost intact, since the right-handed neutrinos decouple at 
sufficiently high scales. Therefore, we ignore the effects from the right-handed neutrinos and the spectrum is effectively the MSSM under the GUT scale in our analyses.

In $4$-$2$-$2$ models, the minimalist choice is to take a universal 
mass parameter $m_{0}$ for all sfermion (sleptons and squarks) 
SSB mass terms at the GUT scale. 
However, the SUSY corrections to the Higgs mass require heavy
($O\textrm{(TeV)}$) third family sfermions 
which would inevitably lead to a heavy SUSY spectrum 
for even the first two family sfermions. 
Motivated with focusing on LHC accessible SUSY solutions,
we consider split soft supersymmetry breaking (SSB) mass term assignment 
parameters, with $m_{12}$ the common mass term for
the first two families' sfermions and $m_{3}$ for the third
family sfermions. Such an assignment can be motivated 
theoretically using flavor symmetries like $SU(2)_f$
\cite{Babu:2014lwa,Babu:2014sga,Hussain:2017fbp}.

The choice of working with a left-right symmetric $4$-$2$-$2$ model, 
relates the gaugino masses at the GUT scale, via
\begin{equation}
	M_1 = \frac{3}{5}M_2 + \frac{2}{5}M_3, 
\end{equation}
and so leads to a more constrained model. This, again is done to allow for 
some freedom in the gaugino sector, but doing so in a minimalist way. 
The sign of the bi-linear Higgs parameter 
plays a crucial role from the point of view of Yukawa coupling unification 
with ${\rm sgn} (\mu) =-1$ being preferred\cite{Gomez:2003cu,Gogoladze:2011ce}. On the other hand, the SUSY
contributions to $\Delta a_\mu$ are proportional to $\mu M_2$.  These considerations 
motivate our choice of $\mu, M_2 < 0$. The so-called $4$-$2$-$2$ inspired MSSM parameter 
space that we scan is given in Table~\ref{table1}.
\begin{center}
\begin{tabular}{c|c|c}
	\hline
	Parameter(s) & Min Value & Max Value \\
	\hline\hline
	$m_{12}$ & 0 & $1.5\,{\rm TeV}$ \\
	$m_3$, $m_{Hu}$, $m_{Hd}$ & 0 & $5\,{\rm TeV}$ \\
	$M_2$ & $-2\,{\rm TeV}$ & 0 \\
	$M_3$ &  0  &$5\,{\rm TeV}$ \\
	$A_0/m_3$ &  -3  &3\\
	$\tan\beta$ & 0 & 60\\
	\hline
\end{tabular}
\label{table1}
\end{center}

We have used the MATHEMATICA package SARAH \cite{Staub:2008uz,Staub:2015iza} which generates the source code for the particle spectrum generator SPheno \cite{Porod:2003um,Goodsell:2014bna}, and MicrOMEGAs \cite{Belanger:2018ccd}, which calculates the DM observables. We have employed a modified Metropolis-Hastings algorithm to scan the relevant parameter space. In our scanning process, we have only retained solutions for which the neutralino is the lightest supersymmetric particle (LSP) to accommodate the DM implications.

Branching ratios of rare $B$ processes ($b \to s \gamma$, $B_{s} \to \mu^{+} \mu^{-}$ and $B_{u} \to \tau \nu_{\tau}$) have been used to constrain the parameter space of new physics models~\cite{Ellis:2007fu,Altmannshofer:2009ne,Bartl:2001ka, Arbey:2012ax,Mahmoudi:2012uk,Mahmoudi:2012un,Hussain:2017fbp,Babu:2014sga, Babu:2014lwa}.
We have applied constraints from branching fractions of the rare decays 
$B \to X_s \gamma$~\cite{HeavyFlavorAveragingGroupHFLAV:2024ctg}, $B_s \to \mu^+ \mu^-$~\cite{CMS:2022mgd,LHCb:2021vsc,LHCb:2021vsc}, $B_u \rightarrow \tau \nu_{\tau}$~\cite{Belle:2012egh,Belle:2015odw} given as,
\begin{align}\label{BPhysVals}
BR(B \rightarrow X_s \gamma) =  (3.49 \pm 0.19 )\times 10^{-4},  \nonumber \\ 
BR(B_s \rightarrow \mu^{+}\mu^{-}) =  (3.34 \pm 0.27)\times 10^{-9} \nonumber\\
BR(B_u \rightarrow \tau \nu_{\tau}) =  (1.09 \pm 0.24 )\times 10^{-4}. 
\end{align}
The latest bounds on the masses of the smuon, stau, chargino, and neutralino are reported by the ATLAS and CMS collaborations as 
exclusion limits in 2D neutralino-smuon, neutralino-stau, and neutralino-chargino planes~\cite{ATLAS:2019lff,ATLAS:2024fub,CMS:2021cox}. 
We have applied the mass bounds on the supersymmetric particles as 
arising from direct searches at the ATLAS and CMS experiments at the 95\% 
exclusion limit. The details are given in Appendix~A.
Further, we impose the following mass bounds
\begin{align}\label{sparticles}
	m_{\tilde{\chi}_1^0} \gtrsim 100 \, \text{GeV} \text{~\cite{ALEPH:2002gap}},\nonumber \\
	122 \,{\rm GeV}   \lesssim m_{h} \lesssim 128 \,\text{GeV}\text{~\cite{ATLAS:2023oaq,CMS:2020xrn,ATLAS:2015yey}}. 
\end{align}
Additionally, we report regions of the parameter space which satisfy the upper bound on the cosmological abundance of a dark matter relic, 
which in our case is the neutralino, \emph{i.e.} $\Omega h^2 \lesssim 0.12$~\cite{Planck:2018vyg}. We do not consider the lower bound due to two reasons: (1) the possibility of multiple sources for dark matter, and (2) a low dark matter relic density ($\Omega h^2 \lesssim 0.12$) in the MSSM  is very finely tuned 
in the case of bino-like solutions which rely on some sort of resonant
annihilation/coannihilation mechanism to cure the relic density. In these cases,
one can almost be certain of finding a point in the neighborhood with a `correct' relic density if we consider only the upper bound. We will later discuss that we have 
bino-like and wino-like solutions in our scan.

It is important to comment on the recent results announced regarding the 
computation of the muon anomalous magnetic moment $\Delta a_\mu$~\cite{Muong-2:2025xyk,Aliberti:2025beg}. It seems that 
the community is converging to an understanding that the seemingly alarming anomaly~\cite{Muong-2:2021ojo,Aoyama:2020ynm} in this sector 
has disappeared and now the SM result agrees with the experimental one within $3\sigma$. 
In so far as this study is concerned, we are using the current $2 \sigma$  upper bound ($0\lesssim\Delta a_\mu\lesssim 6.75\times 10^{-10}$) to constrain our available parameter space. In so doing, we ensure that we are doing no worse than the SM on the side of the lower bound on $\Delta a_\mu$.

\section{Results}\label{Results}

\begin{figure}[H]
	\centering
	\includegraphics[width=1.00\linewidth]{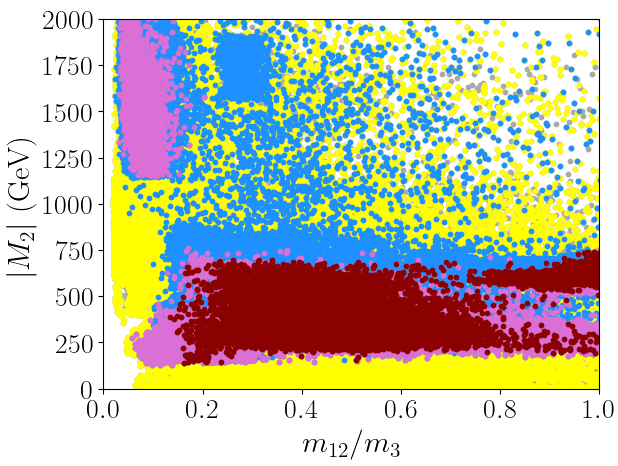}  
	\caption{The $\left(m_{12}/m_3, M_2\right)$ plane. The constraints are shown in different colors and applied successively in the order: data consistent with neutralino LSP (grey), B Physics constraints (yellow), mass bounds from ATLAS and CMS (blue), cosmological abundance of dark matter (orchid) and points satisfying $\Delta a_\mu$ bounds within the $2 \sigma$ range (red). We emphasize that the constraints are applied successively, so that red points satisfy all 
		constraints considered while orchid points satisfy all constraints other than the bounds coming from $\Delta a_\mu$. 
		\label{m12Overm3-M2}}
\end{figure}

We begin the discussion of our results by looking at the behavior of the 
model parameters given experimental constraints. 
In Fig.~\ref{m12Overm3-M2} we discuss results in the $\left(m_{12}/m_3, M_2\right)$ plane and show the effect of applying various constraints successively. 
In grey we show all the data that was collected (consistent with neutralino LSP),
while the yellow points are a subset of the grey points and satisfy the
constraints from $B$ physics given in Eq.~(\ref{BPhysVals}). The blue points 
form a subset of yellow points that additionally satisfy the various bounds on
masses of particles provided by the ATLAS and the CMS experiments as explained 
in Section~\ref{sec2}. 
In orchid we show the subset of blue points that, in addition, satisfy the bounds 
on cosmological abundance of dark matter. Finally, the dark red points comprise 
the subset of orchid points that is consistent with the muon $\Delta a_\mu$ 
result as discussed in Section~\ref{sec2}, in addition to all the other constraints. 

The first thing to note about Fig.~\ref{m12Overm3-M2} is the impact 
of the particle mass bounds shown in blue preferring solutions with $m_{12}/m_3\lesssim 0.5$. This is indicative of the stringent 
constraint on the gluino mass which benefits from a 
large correction from $m_3$. The yellow points in the region of small $m_{12}/m_3$ can clearly be seen to be excluded by the mass bounds for 
$\lvert M_2\rvert\lesssim 900\,{\rm GeV}$. The reason for this is the ATLAS bounds in the chargino-neutralino plane.  
Furthermore, the points consistent with dark matter relic density (orchid points) 
prefer the lower range of $\lvert M_2\rvert\lesssim 750 \, \textrm{GeV}$. Here the main channel for efficient dark matter annihilation in our parameter space is bino-wino coannihilation. In the case of dark matter consistent region 
for larger $\lvert M_2 \rvert$ in this figure, the dominant mechanism is 
selectron (and smuon which is nearly degenerate with selectron) coannihilation which is facilitated by a small 
$m_{12}/m_3$. It is to be noted 
that the $800~\textrm{GeV} \lesssim \lvert M2\rvert\lesssim 1100 ~\textrm{GeV}$ range 
is being excluded by the mass bounds in the neutralino-chargino plane which exclude intermediate 
mass charginos~\cite{Aad:2015eda}.

We also observe from Fig.~\ref{m12Overm3-M2} that the region of parameter space allowed by the $\Delta a_\mu$ constraint (points in red) favors either a small $\lvert M_2\rvert$ (corresponding to light wino) or a small $m_{12}/m_3$ that allows for the possibility of a light muon-sneutrino ($\tilde\nu_{\mu}$).  

\begin{figure}[t!]
	\centering
	\includegraphics[width=1.00\linewidth]{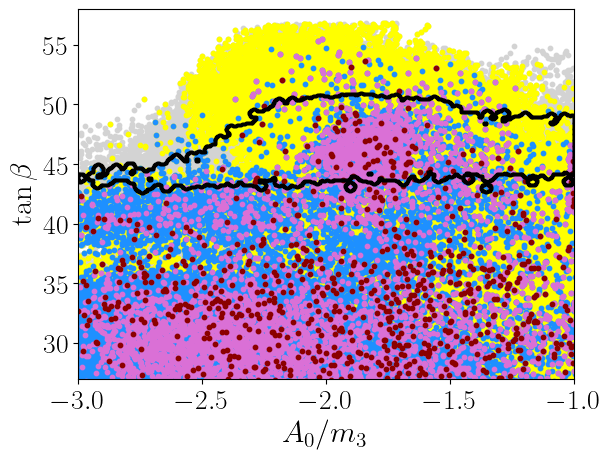}  
	\caption{The $\left(A_{0}/m_3, \tan\beta\right)$ plane. Color coding is the same as in Fig.~\ref{m12Overm3-M2}. 
		Further we show the contour (in black) corresponding to $R_{tb\tau}\simeq 1.1.$
		\label{A0Overm3-tanb}}
\end{figure}
In Fig.~\ref{A0Overm3-tanb} we show results in the $\left(A_{0}/m_3, \tan\beta\right)$ plane. We follow the same color coding as we did for Fig.~\ref{m12Overm3-M2}. In Fig.~\ref{A0Overm3-tanb} we have also drawn a contour corresponding to $R_{tb\tau}\simeq 1.1$, where
\[R_{tb\tau}\equiv\frac{\textrm{max}(y_t,y_b,y_\tau) (M_{\textrm GUT})}{{\rm min}(y_t,y_b,y_\tau) (M_{\textrm GUT})}\]
is the ratio of the maximum of the $t,b,\tau$ Yukawa couplings to the minimum of them and provides an appropriate parameter to quantify Yukawa coupling unification.
We can see from Fig.~\ref{A0Overm3-tanb} that Yukawa 
coupling unification favors large $\tan\beta$ solutions
with $43\lesssim \tan\beta \lesssim 52$, as expected. We
can see from the density of points in blue, orchid and dark-red
falling into the $R_{tb\tau}\simeq 1.1$ contour that we
should expect Yukawa unified solutions consistent with all
the experimental constraints that we consider. While 
Fig.~\ref{A0Overm3-tanb} is suggestive, we can 
confirm that no particular value of $A_0/m_3<0$ is needed
for Yukawa unification. In particular, while it 
is not apparent from this figure, one can 
accommodate solutions with even
$R_{tb\tau}\to 1$ can be realized in our model\cite{Gogoladze:2010fu,Gogoladze:2011ce}. In addition, it is also obvious from the
contour that a broad range of $A_0/m_3$ values provide
Yukawa unified solutions consistent with all experimental constraints. 
\begin{figure}[H]
	\centering
	\includegraphics[width=1.00\linewidth]{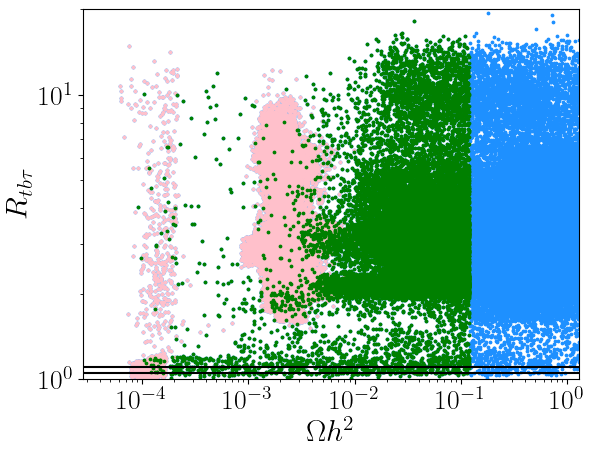}  
	\caption{The $(\Omega h^{2}, R_{tb\tau})$ plane. The blue points satisfy B Physics constraints and mass bounds as in the previous figures. The pink~(green) points form that subset of blue 
points that are predominantly wino~(bino) in their constitution 
and have $\Omega h^2\lesssim 0.12$. The horizontal lines correspond to $R_{tb\tau}=1.05,\, 1.10.$
		\label{omegaR}}
\end{figure}
In Fig.~\ref{omegaR} we focus on the type of the dark matter 
solutions that we have found in our scan. We show the results 
in the $(\Omega h^2,R_{tb\tau})$ plane with points in blue 
satisfying the B Physics and mass bounds as in the case of the previous figures. The pink~(green) points form that subset of blue 
points for which $\Omega h^2\lesssim 0.12$ and which are predominantly wino~(bino) in their constitution with the 
wino~(bino) component larger than $90\%$. We refer here to Baer \emph{et al} 
who show that in cases with opposite sign gauginos there is no continuous 
change between a bino-like and wino-like neutralino and this 
happens rather abruptly \cite{Baer:2005jq}. We have plotted the green points after the 
pink so some of them are hidden underneath. We have also 
drawn two horizontal lines in this plane corresponding 
to $R_{tb\tau}=1.05,\,1.10$ signifying the region of $5\%$ and 
$10\%$ or better Yukawa coupling unification. As explained earlier, 
we can see solutions with $R_{tb\tau}\to 1$ in the 
case of both wino-like and bino-link dark matter consistent with 
all other constraints. 
\begin{figure}[t!]
	\centering
	\includegraphics[width=1.00\linewidth]{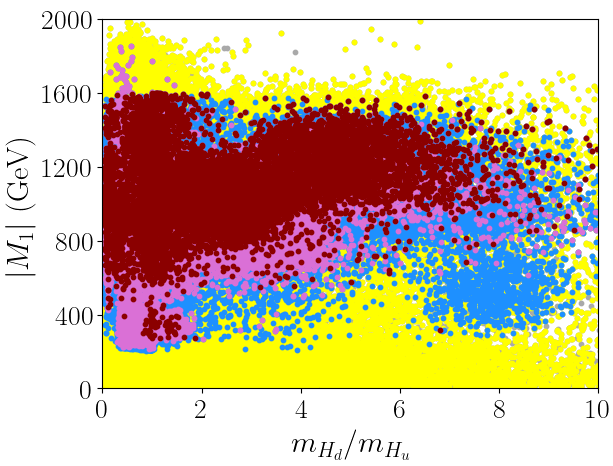}  
	\caption{The $\left( m_{H_d}/m_{H_u}, M_{1}\right)$ plane. Color coding is the same as in Fig.~\ref{m12Overm3-M2}. } 
		\label{M1-mHdmHu}
\end{figure}

In Fig.~\ref{M1-mHdmHu} we present our results in
the ($m_{H_d}/m_{H_u}, M_1$) plane. We can infer from this figure that the dark matter relic density favors $\lvert M_1\rvert \gtrsim 200\,{\rm GeV}$ but we want to emphasize on the fact that points with $M_1 \lesssim 200~\textrm{GeV}$ do have the correct relic density (due to, for instance Z-pole and SM-like Higgs pole region) but are excluded by mass bounds on various particles.

We can see a predictable behavior in Fig.~\ref{M1-mHdmHu} from the point of view of $\Delta a_\mu$ wherein $M_1\lesssim 1.5\,{\rm TeV}$ is preferred. This is to be expected as $M_1$ 
is correlated with the LSP mass and so sets a scale for any 
SUSY contribution to SM processes.

\begin{figure}[b!]
	\centering
	\includegraphics[width=1.00\linewidth]{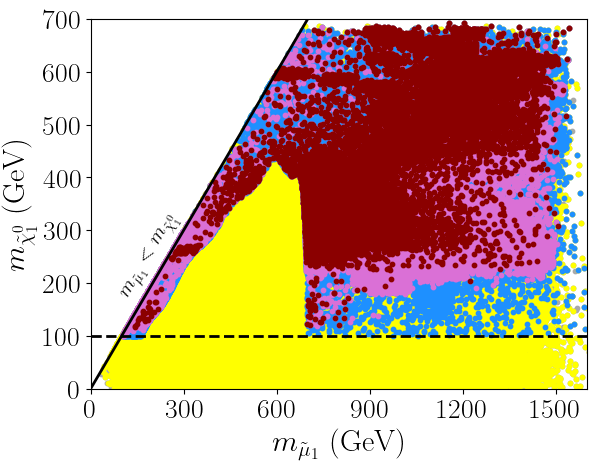}  
	\caption{The $\left(  m_{\tilde{\mu}_{1}}, m_{\tilde{\chi}^{o}_{1}} \right)$ plane. Color coding is the same as in Fig.~\ref{m12Overm3-M2}.}
	\label{smuon1Neut}
\end{figure}

We now turn our attention to the various dark matter allowed 
scenarios that can be found in our model and are 
consistent with the various experimental constraints we have discussed. In Fig.~\ref{smuon1Neut} we show the smuon coannihilation region (orchid and red) next to the line $m_{\tilde{\chi}_1^0} = m_{\tilde{\mu}_1}$. The selectron and the smuon in our model are effectively degenerate, 
since their masses are determined by $m_{12}$ and do not split much via the renormalization group equations (RGEs). Thus the solutions in this plane also 
indicate selectron coannihilation.
The peculiar 
triangular shape of the yellow region is due to the 
ATLAS constraints in the $(m_{\tilde{\mu}_1}, m_{\tilde{\chi}_1^0})$ plane with the blue, orchid and red points satisfying these constraints. 
The blue band sans orchid in the roughly rectangular region 
$ m_{\tilde{\mu}_1}/{\rm GeV}\gtrsim 700$ and 
$160 \lesssim m_{\tilde{\chi}_1^0}/{\rm GeV}\lesssim 180$ 
arises as chargino coannihilation is the main 
mechanism of curing the dark matter relic density in this 
neighborhood. This region is naturally sensitive to the 
ATLAS bounds in the $(m_{\tilde{\chi}_1^\pm}, m_{\tilde{\chi}_1^0})$ plane, and we did not find consistent solutions 
in our scan. This is likely a numerical artifact of our simulation. 
\begin{figure}[H]
	\centering
	\includegraphics[width=1.00\linewidth]{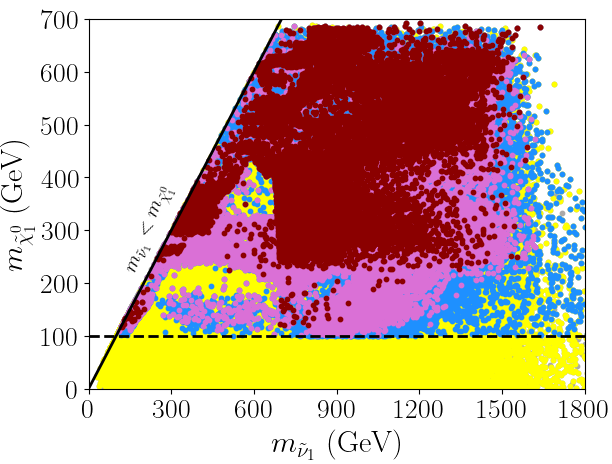}  
	\caption{The $\left(  m_{\tilde{\nu}_{1}}, m_{\tilde{\chi}^{o}_{1}}  \right)$ plane. Color coding is the same as in Fig.~\ref{m12Overm3-M2}.} 
	\label{snu1Neut}
\end{figure}  

\begin{figure}[H]
	\centering
	\includegraphics[width=1.00\linewidth]{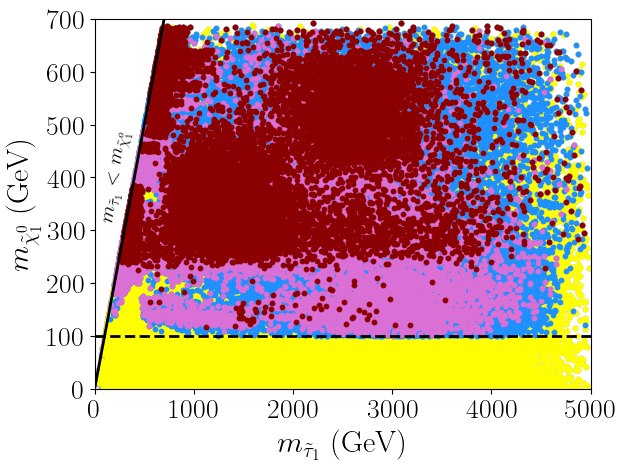}  
	\caption{The $\left(  m_{\tilde{\tau}_{1}}, m_{\tilde{\chi}^{o}_{1}}  \right)$ plane. Color coding is the same as in Fig.~\ref{m12Overm3-M2}.}
	\label{stau1Neut}
\end{figure}

We see the general features seen in Fig.~\ref{smuon1Neut} 
repeat in Fig.~\ref{snu1Neut} where we now shift focus to the $\left(  m_{\tilde{\nu}_{1}}, m_{\tilde{\chi}^{o}_{1}}  \right)$ plane. We can see the sneutrino-coannihilation region along the line $m_{\tilde{\chi}^{o}_{1}} = m_{\tilde{\nu}_{1}}$. In this plane, the typical channel for efficient neutralino annihilation mostly 
involves the chargino, sneutrino and the selectron/smuon for 
$m_{\tilde{\nu}_{1}} \lesssim 1500\,{\rm GeV}$, with the 
sneutrino contribution significant only when it is the NLSP. 
Similarly, in Fig.~\ref{stau1Neut} we can clearly see 
the stau-coannihilation region with a 
neutralino as light as $\sim 200\,{\rm GeV}$. 
\begin{figure}[H]
	\centering
	\includegraphics[width=1.00\linewidth]{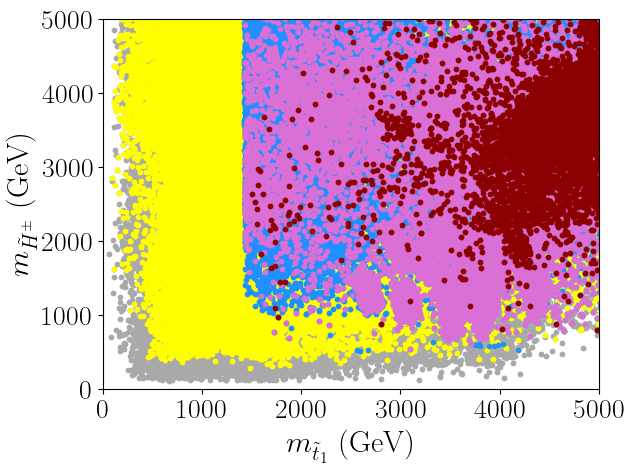}  
	\caption{The $\left(  m_{\tilde{t}_1}, m_{H^{\pm}}  \right)$ plane. Color coding is the same as in Fig.~\ref{m12Overm3-M2}.}
	\label{stop1Hpm}
\end{figure}
In Fig.~\ref{stop1Hpm}, we present our results in the $(m_{\tilde{t}_1},m_{H^{\pm}})$ 
plane. As the decay rates of rare B meson receive SUSY contributions from Higgs sector and stop, we can see that the constraints from the branching ratios of the B decays 
given in Section~\ref{sec2} put lower mass bounds on both $m_{H^{\pm}}$ and $m_{\tilde{t}_1}$ that can be seen as the 
visible grey regions where these masses are `small'. 
\begin{figure}[H]
	\centering
	\includegraphics[width=1.00\linewidth]{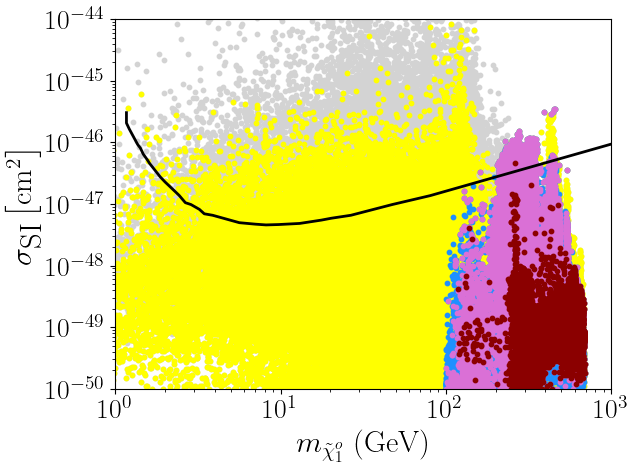}  
	\caption{The $\left(m_{\tilde{\chi}^{o}_{1}}, \sigma_{SI}\right)$ plane. Color coding is the same as in Fig.~\ref{m12Overm3-M2}. The black line shows the latest bounds on WIMP-nucleon spin independent cross-section by the LUX-ZEPLIN Collaboration\cite{LZ:2022lsv}.} 
	\label{sigmaSI}
\end{figure}

We now focus our attention on ground based dark matter  detection experiments that put limits on the the spin-independent ($\sigma_{SI}$) and spin-dependent ($\sigma_{SD}$) nucleon-WIMP cross-sections. 
Currently the most stringent bounds on $\sigma_{SI}$ and 
$\sigma_{SD}$ (neutron) are 
reported by the LUX-ZEPLIN collaboration while PICO is competitive in the case of $\sigma_{SD}$ (proton) case. In Fig.~\ref{sigmaSI} we plot the 
$\sigma_{SI}$-nucleon cross sections as a function of the neutralino mass. We show, conservatively, the $90 \%$ exclusion limit incorporating the $2\sigma$ uncertainty in the experimental result as a
solid line in Fig.~\ref{sigmaSI}.  
It can be seen that some pink points (satisfying all bounds except the one from $\Delta a_{\mu}$) are excluded by this 
result. More interestingly, this experiment should be 
applying increasingly stringent bounds on the parameter space in its current run. We have also looked at $\sigma_{SD}$ which shows a similar trend with the current experimental limit just at the edge of the otherwise allowed region. 

\section{Conclusion}\label{conclusion}
In this work, we have explored the phenomenological viability of a GUT-inspired MSSM 
framework based on a left-right symmetric $SU(4)c \times SU(2)_L \times SU(2)_R$ model with 
non-universal soft SUSY-breaking terms. \rt{During the symmetry breaking all the additional fields acquire heavy masses comparable with $M_{{\rm GUT}}$ such that only the MSSM fields and their interactions remain available below $M_{{\rm GUT}}$}. By employing a detailed numerical scan of the 
parameter space using the SARAH and SPheno toolchains, we assessed the model against a 
comprehensive set of experimental constraints, including those from B-physics observables, 
dark matter relic density, direct detection limits, and collider searches from ATLAS and CMS.

Our analysis identifies viable regions of parameter space that satisfy the requirement of 
third-generation Yukawa coupling unification and remain consistent with the $2\sigma$ bounds 
on the anomalous magnetic moment of the muon ($\Delta a_\mu$). These regions are 
characterized by either slepton or bino-wino coannihilation mechanisms that efficiently 
deplete the neutralino dark matter relic density to the observed value.

While current LHC bounds significantly restrict the low-energy spectrum, our results demonstrate that well-motivated and testable MSSM scenarios persist. These models offer promising avenues for further exploration at ongoing and upcoming collider runs and dark matter experiments such as LUX-ZEPLIN. Future experimental advances will continue to play a decisive role in probing or constraining this class of supersymmetric theories.

\section{Acknowledgment}
The work of CSU is supported by the Scientific and Technological Research Council of Türkiye (TUBITAK) Grant. No MFAG-125F122. The numerical
calculations reported in this paper were partially performed at TUBITAK ULAKBIM, High Performance and Grid Computing Center (TRUBA resources). We also acknowledge the LUMS High Performance Computing Centre (HPCC)\footnote{{\url{https://hpc.lums.edu.pk}}} (managed by Nouman Zubair) for providing the computing resources and technical support for this study under Allocation ID: LUMS-HPC-2025-01 for part of this study.
{\onecolumngrid
\begin{center}
\rule{0.5\textwidth}{0.8pt} 
\end{center}
}
\twocolumngrid
\appendix*
\section{Mass bounds on the spectrum}
\begin{figure}[H]
	\centering
	\includegraphics[width=.80\linewidth]{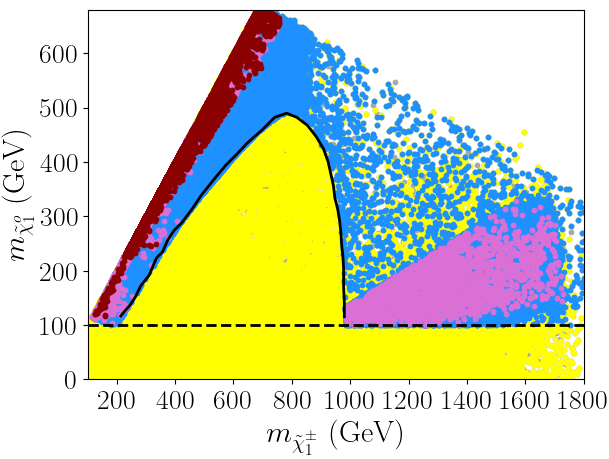}  
	\caption{The $ \left( m_{\tilde{\chi}^{\pm}_{1}},m_{\tilde{\chi}^{0}_{1}} \right)$ plane. The black curve represents the 95$\%$ exclusion limit in this plane as given in \cite{ATLAS:2024fub,ATLAS:2019lff}. The horizontal dashed line represents the LEP bound on $m_{\tilde{\chi}^{0}_{1}}$~\cite{ALEPH:2002gap}. The rest of the color coding is the same as in Fig.~\ref{m12Overm3-M2}.} 
	\label{chNe}
\end{figure}

\begin{figure}[H]
	\centering
	\includegraphics[width=.80\linewidth]{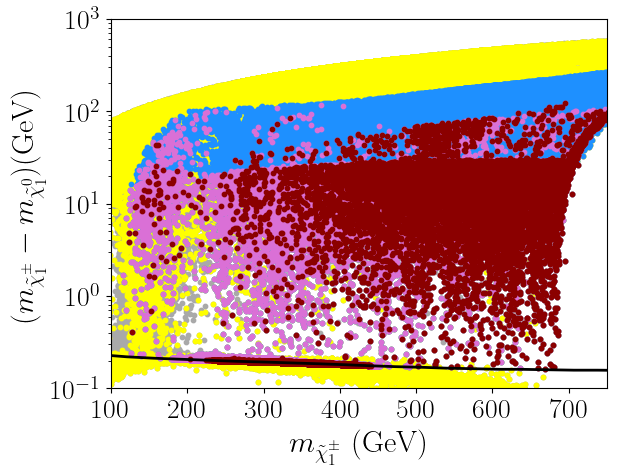}  
	\caption{The $\left(m_{\tilde{\chi}^{\pm}_{1}} , \left(m_{\tilde{\chi}^{\pm}_{1}} -m_{\tilde{\chi}^{0}_{1}}\right)\right)$ plane. The black curve represents the 95$\%$ exclusion limit in this plane as given in \cite{CMS:2023mny}. The rest of the color coding is the same as in Fig.~\ref{chNe}.} 
	\label{chDl}
    \end{figure}

\begin{figure}[H]
	\centering
	\includegraphics[width=.80\linewidth]{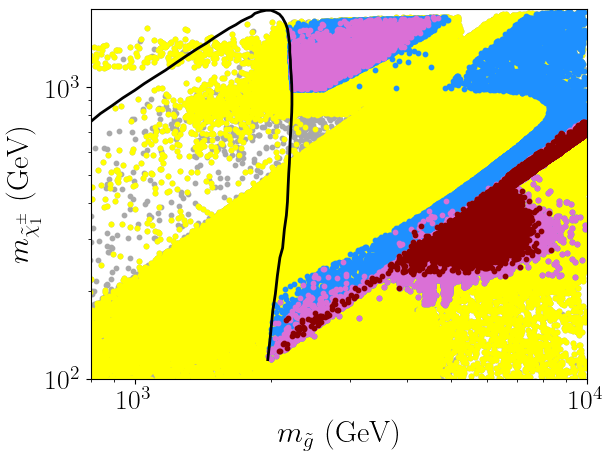}  
	\caption{The $\left(m_{\tilde{g}} , m_{\tilde{\chi}^{\pm}_{1}}\right)$ plane. The black curve represents the 95$\%$ exclusion limit in this plane as given in \cite{ATLAS:2022rme}. The rest of the color coding is the same as in Fig.~\ref{chNe}.} 
	\label{glCh}
\end{figure}

\begin{figure}[H]
	\centering
	\includegraphics[width=.80\linewidth]{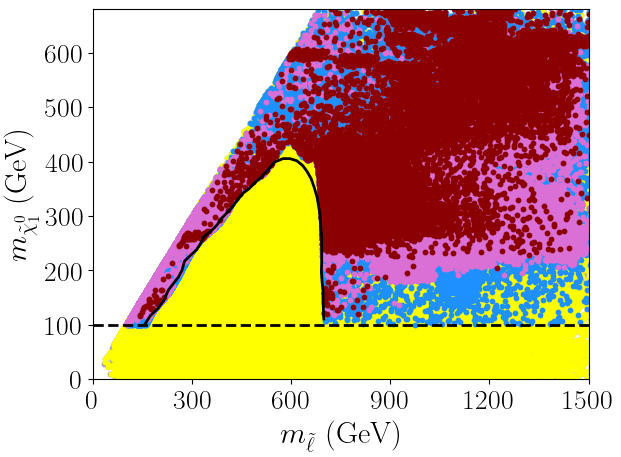}  
	\caption{The $\left(m_{\tilde{\ell}} , m_{\tilde{\chi}^{0}_{1}}\right)$ plane. The black curve represents the 95$\%$ exclusion limit in this plane as given in \cite{ATLAS:2019lff}. The rest of the color coding is the same as in Fig.~\ref{chNe}.} 
	\label{slNe}
\end{figure}

\begin{figure}[H]
	\centering
	\includegraphics[width=.80\linewidth]{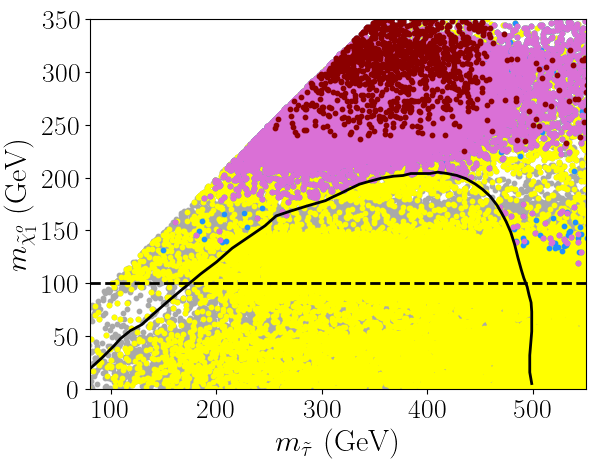}  
	\caption{The $\left(m_{\tilde{\tau}} , m_{\tilde{\chi}^{\pm}_{1}}\right)$ plane. The black curve represents the 95$\%$ exclusion limit in this plane as given in \cite{ATLAS:2024fub}. The rest of the color coding is the same as in Fig.~\ref{chNe}. } 
	\label{stauNe}
\end{figure}

\begin{figure}[H]
	\centering
	\includegraphics[width=.80\linewidth]{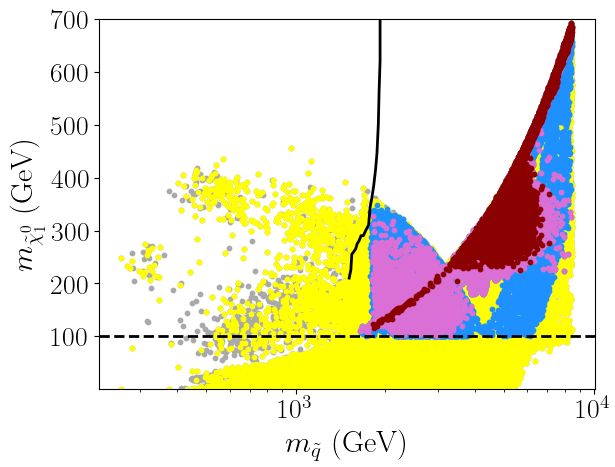}  
	\caption{The $\left(m_{\tilde{q}} , m_{\tilde{\chi}^{0}_{1}}\right)$ plane. The black curve represents the 95$\%$ exclusion limit in this plane as given in \cite{CMS:2023zuu}. The rest of the color coding is the same as in Fig.~\ref{chNe}.} 
	\label{sqNe}
\end{figure}

\begin{figure}[H]
	\centering
	\includegraphics[width=.80\linewidth]{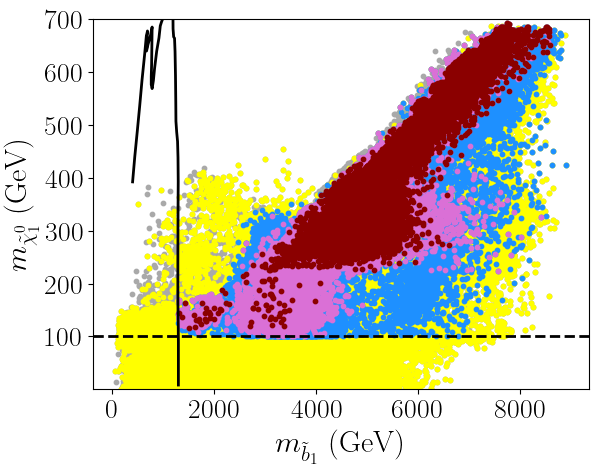}  
	\caption{The $\left(m_{\tilde{b}_1} , m_{\tilde{\chi}^{0}_{1}}\right)$ plane. The black curve represents the 95$\%$ exclusion limit in this plane as given in \cite{ATLAS:2021yij}. The rest of the color coding is the same as in Fig.~\ref{chNe}.} 
	\label{sbNe}
\end{figure}

\begin{figure}[H]
	\centering
	\includegraphics[width=.80\linewidth]{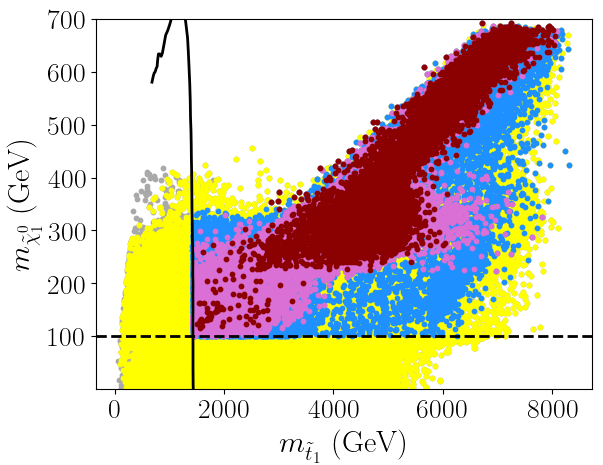}  
	\caption{The $\left(m_{\tilde{t}_1} , m_{\tilde{\chi}^{0}_{1}}\right)$ plane. The black curve represents the 95$\%$ exclusion limit in this plane as given in \cite{CMS:2021beq}. The rest of the color coding is the same as in Fig.~\ref{chNe}.} 
	\label{stopNe}
\end{figure}

We now show the plots in the $( m_{\tilde{\chi}^{\pm}_{1}} , m_{\tilde{\chi}^{0}_{1}})$,
$(m_{\tilde{\chi}^{\pm}_{1}} , \Delta m)$, where $\Delta m = m_{\tilde{\chi}^{\pm}_{1}}-m_{\tilde{\chi}^{0}_{1}} $, 
$(m_{\tilde{g}} , m_{\tilde{\chi}^{\pm}_{1}})$, 
$(m_{\tilde{\ell}} ,  m_{\tilde{\chi}^{0}_{1}} )$, 
$(m_{\tilde{\tau}} , m_{\tilde{\chi}^{0}_{1}})$, 
$(m_{\tilde{q}} ,  m_{\tilde{\chi}^{0}_{1}})$,
$(m_{\tilde{b}} , m_{\tilde{\chi}^{0}_{1}})$, 
and
$(m_{\tilde{t}} , m_{\tilde{\chi}^{0}_{1}})$ 
planes, in Fig.~\ref{chNe} to Fig.~\ref{stopNe}, where we have applied the mass bounds as given by the latest ATLAS, and CMS results. The color coding is the same as in Fig.~\ref{m12Overm3-M2}. In all of these plots, we can 
see the allowed regions from the point of view of the mass bounds in blue, orchid 
and red (with orchid and red being subsets of blue). The exclusion 
limits are taken directly from the published literature and correspond to the 
95\% exclusion regions in the corresponding planes. 
\bibliographystyle{ieeetr}
\bibliography{hku-PS-Rev}

\end{document}